\documentclass[11pt]{article}

\usepackage{cite}
\usepackage{url}
\usepackage{amsmath,amssymb,amsfonts}
\usepackage{epsfig}
\usepackage{color}
\usepackage{booktabs}

\usepackage{graphics}
\usepackage{graphicx}
\usepackage{subfigure}

\usepackage[normalem]{ulem}

\usepackage{multirow}

\usepackage{array}

\usepackage[dvips,letterpaper]{geometry}

\usepackage{srcltx}

\usepackage{amsmath}
\usepackage{amssymb,amsfonts}

\usepackage{graphics}
\usepackage{epsfig}
\usepackage{graphicx}
\usepackage{color}
\usepackage[usenames,dvipsnames]{xcolor}

\usepackage{booktabs}

\usepackage{fullpage}
\usepackage{setspace}
\usepackage{flushend}
\usepackage{multicol}
\usepackage{enumerate}

\usepackage{cite}
\usepackage{url}

\usepackage[short,12hr]{datetime}

\usepackage{hyperref}
\usepackage{algorithm}
\usepackage{algpseudocode}
\usepackage{multicol}
       \usepackage{mathptmx}

\makeatletter

\newenvironment{rsmallmatrix}{\null\,\vcenter\bgroup
  \Let@\restore@math@cr\default@tag
  \baselineskip6\ex@ \lineskip1.5\ex@ \lineskiplimit\lineskip
  \ialign\bgroup\hfil$\m@th\scriptstyle##$&&\thickspace\hfil
  $\m@th\scriptstyle##$\crcr
}{%
  \crcr\egroup\egroup\,%
}
\makeatother

\title{A Discrete Tchebichef Transform Approximation
for Image and Video Coding}

\author{%
Paulo A. M. Oliveira$^\ast$
\quad
Renato~J.~Cintra%
\thanks{Paulo A. M. Oliveira and Renato~J.~Cintra 
are with the
Signal Processing Group,
Departamento de Estat\'{\i}stica,
Universidade Federal de Pernambuco, 
Recife, PE, Brazil.
R. J. Cintra is also
with the LIRIS, 
Institut National des Sciences Appliqu\'ees (INSA),
Lyon, France
(e-mail: rjdsc@ieee.org).}
\quad
F\'abio~M.~Bayer%
\thanks{%
F\'abio~M.~Bayer
is with the
Departamento de Estat\'{\i}stica 
and LACESM,
Universidade Federal de Santa Maria, 
Santa Maria, RS, Brazil
(e-mail: bayer@ufsm.br).}
\\
Sunera~Kulasekera$^\ddagger$
\quad
Arjuna~Madanayake%
\thanks{%
Sunera Kulasekera
and
Arjuna Madanayake
are with the
Department of Electrical and Computer Engineering,
The University of Akron, Akron, OH, USA
(e-mail: arjuna@uakron.edu).}
}

\date{}

\begin{document}

\maketitle

\onehalfspacing

\begin{abstract} 
In this paper,
we introduce a low-complexity approximation
for the discrete Tchebichef transform (DTT).
The proposed forward and inverse transforms are multiplication-free
and require a reduced number of additions and bit-shifting operations.
Numerical compression simulations
demonstrate the efficiency of the proposed transform
for image and video coding.
Furthermore, 
Xilinx Virtex-6 FPGA based hardware realization shows 44.9\% reduction in
dynamic power consumption and 64.7\% lower area when compared to the
literature.
\end{abstract}

\begin{center}
\small
\textbf{Keywords}
\\
Approximate DTT,
fast algorithms, image and video coding.
\end{center}

\section{Introduction}

The discrete Tchebichef transform (DTT)
is a 
useful 
tool for
signal coding and data decorrelation~\cite{Mukundan2001DTM}.
In recent years,
signal processing literature
has employed the DTT
in several image processing problems,
such as
artifact measurement~\cite{Leida2014artifact_tchebichef},
blind integrity verification~\cite{Roux2012blind},
and
image compression~\cite{Ernawan2013quantization_tchebichef,Pratipati2013DTT_quant,Swamy2013ITT,Senapati2014listlessDTT}.
In particular,
the 8-point DTT
has been considered in blind forensics
for integrity check of medical images~\cite{Roux2012blind}.
For image compression,
the 8-point DTT 
is
also
capable of outperforming
the 8-point discrete cosine transform (DCT)
in terms of average bit-length in bitstream codification~\cite{Ernawan2013quantization_tchebichef}.
Moreover, 
in~\cite{Senapati2014listlessDTT}
an 8-point DTT-based
encoder 
capable of 
improved image quality and 
reduced encoding/decoding time
was proposed;
being a competitor to
state-of-the-art DCT-based methods.
However, 
to the best of our knowledge, 
literature archives only one fast algorithm for the 8-point DTT,
which requires 
a significant number of arithmetic operations~\cite{Swamy2013ITT}.
Such high
arithmetic complexity
may be a hindrance for the adoption of the DTT
in contemporary devices
that
demand
low-complexity circuitry
and
low power consumption~\cite{Li2008WSN,Ernawan2011mobile_tchebichef,Kouadria2013WSN}.

An alternative to the exact transform computation
is the employment of approximate transforms.
Such approach has been successfully applied 
to the exact DCT, 
resulting in several approximations~\cite{CB2011RDCT,  cintra2014dct_aprox}.
In general, 
an approximate transform 
consists of a low-complexity matrix
with elements defined over
a set of small integers,
such as
$\{0,\pm1,\pm2,\pm3\}$.
The resulting matrix possesses
null multiplicative complexity,
because the involved arithmetic operations can be
implemented exclusively by means of 
a reduced number of additions and bit-shifts.
Prominent examples of approximate transforms
include:
the signed DCT~\cite{Haweel2001SDCT},
the series of DCT approximations by Bouguezel-Ahmed-Swamy~\cite{BAS2008,BAS2011,BAS2013},
the approximation by Lengwehasatit-Ortega~\cite{LO2004Dct_appr},
and
the integer based approximations described in~\cite{CB2011RDCT,CB2012MRDCT,cintra2014dct_aprox,Potluri2014Improved_Approx}.

In this work, 
we introduce a low-complexity DTT approximation 
that
requires 54.5\%~less additions than the exact DTT fast algorithm.
The proposed method
is suitable for image and video coding, 
capable of processing
data
coded
according to 
popular standards---such as 
JPEG~\cite{Wallace1992JPEG},
H.264~\cite{H264_book},
and
HEVC~\cite{Sullivan2012HEVC}---at a low computational cost.
Moreover,
the FPGA hardware realization
of the proposed transform
is also sought.

This paper unfolds as follows.
Section~\ref{section-adtt} describes the DTT
and
introduces
the approximate DTT
with its associate fast algorithm.
A computational complexity analysis is offered. 
In Section~\ref{section-experimental}, 
we perform numerical experiments;
applying of the proposed transform 
as a tool for image and video compression. 
In Section~\ref{vlsi}, 
we provide very large scale integration (VLSI)
realizations of the exact DTT and proposed approximation. 
Conclusions and final remarks are in Section~\ref{conclusion}.

\section{Discrete Tchebichef Transform Approximation}
\label{section-adtt}

\subsection{Exact Discrete Tchebichef Transform}

The DTT is an orthogonal transformation
derived 
from the discrete Tchebichef polynomials~\cite{HTF53}.
The entries of the $N$-point DTT matrix
are furnished by~\cite{Mukundan2001DTM}:
\begin{align}
\label{DTT_coeff}
t_{k,n}
=
&
\sqrt{
\frac{(2k+1)(N-k-1)!}{(N+k)!}
}
\cdot
(1-N)_{k}
\cdot
{}_3 F_{2}(-k,-n,1+k;1,1-N;1),
\qquad
k,n = 0, 1, \ldots , N-1
,
\end{align}
where
${}_3 F_{2} (a_1, a_2, a_3; b_1, b_2;z)
=
\sum_{n=0}^{\infty}
\frac{(a_1)_k (a_2)_k (a_3)_k}{(b_1)_k (b_2)_k}
\cdot
\frac{z^k}{k!}
$
is the hypergeometric function
and
${(a)}_k = a(a+1)\cdots(a+k-1)$
is the ascending factorial.
Therefore,
the analysis and synthesis equations
for the DTT are given by
$\mathbf{X}
=
\mathbf{T}
\cdot
\mathbf{x}$
and
$\mathbf{x}
=
\mathbf{T}^{-1}
\cdot
\mathbf{X}
=
\mathbf{T}^\top
\cdot
\mathbf{X}$,
where
$\mathbf{x}= \begin{bmatrix}x_0 & x_1 & \cdots & x_{N-1}\end{bmatrix}^\top$
is the input signal,
$\mathbf{X} = \begin{bmatrix}X_0 & X_1 & \cdots & X_{N-1}\end{bmatrix}^\top$
is the transformed signal,
and
$\mathbf{T}$
is the $N$-point DTT matrix with elements $t_{k,n}$, $k,n=0,1,\ldots,N-1$,

In particular, 
the 8-point DTT matrix~$\mathbf{T}$
can be described by
the product of
a diagonal matrix~$\mathbf{F}$
and an 
integer-entry
matrix~$\mathbf{T}_0$~\cite{Swamy2013ITT},
resulting in:
$
\mathbf{T}
=
\mathbf{F}\cdot\mathbf{T}_0
$,
where
\begin{align}
\mathbf{T}_0
= 
\left[
\begin{rsmallmatrix}
1 & 1 & 1 & 1 & 1 & 1 & 1 & \phantom{-}1 \\
-7 & -5 & -3 & -1 & 1 & 3 & 5 & 7 \\
7 & 1 & -3 & -5 & -5 & -3 & 1 & 7 \\
-7 & 5 & 7 & 3 & -3 & -7 & -5 & 7 \\
7 & -13 & -3 & 9 & 9 & -3 & -13 & 7 \\
-7 & 23 & -17 & -15 & 15 & 17 & -23 & 7 \\
1 & -5 & 9 & -5 & -5 & 9 & -5 & 1 \\
-1 & 7 & -21 & 35 & -35 & 21 & -7 & 1
\end{rsmallmatrix}
\right]
,
\end{align}
and
$\mathbf{F} =
\frac{1}{2}
\cdot
\operatorname{diag}
\left(
\frac{1}{\sqrt{2}},
\frac{1}{\sqrt{42}},
\frac{1}{\sqrt{42}},
\frac{1}{\sqrt{66}},
\frac{1}{\sqrt{154}},
\frac{1}{\sqrt{546}},
\frac{1}{\sqrt{66}},
\frac{1}{\sqrt{858}}
\right)
$.
A fast algorithm for the above integer matrix 
$\mathbf{T}_0 = \mathbf{F}^{-1}\cdot\mathbf{T}$ 
was derived in~\cite{Swamy2013ITT}
requiring
44~additions 
and
29~bit-shifting operations.
Such arithmetic complexity is considered excessive,
when compared to state-of-the-art 
discrete transform approximations
which generally require less than 
24~additions~\cite{cintra2014dct_aprox,LO2004Dct_appr,Haweel2001SDCT,BAS2013}.

\subsection{DTT Approximation and Fast Algorithm}

In~\cite{cintra2014dct_aprox},
a class of DCT approximations
was introduced based on the following relation:
$\operatorname{round}(\alpha \cdot \mathbf{C})$,
where
$\operatorname{round}(\cdot)$
is the round function
as
defined in
C and Matlab languages~\cite{cintra2014dct_aprox},
$\alpha$ is a real parameter,
and
$\mathbf{C}$ is the exact DCT matrix.
We aim at proposing a similar approach
to obtain an 8-point DTT approximation.
The scale-and-round approach
is particularly effective when
discrete trigonometric transforms are considered.
This is because
the entries of such transformation matrices
have
smaller
dynamic ranges
when compared to the DTT.
In contrast,
the DTT entries have values with a dynamic range
roughly seven times larger than the DCT,
for example.
Thus the approximation error implied by
the round function
is less evenly distributed
in non-trigonometric transform matrices,
such as the DTT.
To mitigate this effect,
we propose
a compading-like operation~\cite{Gray1998Quant},
consisting of
a rescaling matrix~$\mathbf{D}$
that
normalizes the DTT matrix entries.
Thus,
according the formalism detailed in~\cite{cintra2014dct_aprox},
we introduce
a parametric family of
approximate DTT matrices~$\mathbf{T}(\alpha)$,
which are
given by:
\begin{align}
\mathbf{T}
(\alpha)
=
\operatorname{round}
\left(
\alpha
\cdot
\mathbf{T}
\cdot
\mathbf{D}_0
\right)
,
\end{align}
where
$\mathbf{D}_0
= 
\operatorname{diag}
(
\sqrt{\frac{6}{7}},
\frac{\sqrt{154}}{13},
\frac{\sqrt{66}}{9},
\frac{\sqrt{858}}{35},
\frac{\sqrt{858}}{35},
\frac{\sqrt{66}}{9},
\frac{\sqrt{154}}{13},
\sqrt{\frac{6}{7}}
)$.

We aim at identifying a particular optimal
parameter~$\alpha^\ast$
such that
$\mathbf{T}^\ast = \mathbf{T}(\alpha^\ast)$
results in a matrix
satisfying the following constraints:
(i)~the entries of $\mathbf{T}^\ast$ 
must be defined over
$\{-1, 0, 1\}$
and
(ii)~$\mathbf{T}^\ast$ must possess low arithmetic complexity.
Constraint~(i)
implies the search space $(0, 3/2)$.
Although
the above problem is not analytically tractable,
its solution can be found by exhaustive search~\cite{cintra2014dct_aprox}.
By taking the values of $\alpha$ over the considered interval 
in steps of $10^{-3}$,
above conditions
are satisfied for
$0.931 \leq \alpha^\ast \leq 0.957$.
All values of $\alpha^\ast$ in this latter interval
imply the same approximate matrix.
Thus,
the obtained low-complexity forward
DTT approximation is given by:
\begin{align}
\mathbf{T}^\ast
=
&
\left[
\begin{rsmallmatrix}
1 & 1 & 1 & 1 & 1 & 1 & 1 & \phantom{-}1 \\
-1 & -1 & 0 & 0 & 0 & 0 & 1 & 1 \\
1 & 0 & 0 & -1 & -1 & 0 & 0 & 1 \\
-1 & 1 & 1 & 0 & 0 & -1 & -1 & 1 \\
0 & -1 & 0 & 1 & 1 & 0 & -1 & 0 \\
0 & 1 & -1 & -1 & 1 & 1 & -1 & 0 \\
0 & -1 & 1 & 0 & 0 & 1 & -1 & 0 \\
0 & 0 & -1 & 1 & -1 & 1 & 0 & 0 \\
\end{rsmallmatrix}
\right]
\end{align}
and its inverse 
$\mathbf{T}^\ast$
is given by:
$
(\mathbf{T}^\ast)^{-1} 
=
\mathbf{T}_1
\cdot
\mathbf{D}_1
$
where
\begin{align}
\mathbf{T}_1
=
&
\left[
\begin{rsmallmatrix}
1 & -3 & 3 & -2 & 1 & -1 & -1 & -1 \\
1 & -2 & -1 & 2 & -1 & 1 & -1 & 1 \\
1 & -1 & -1 & 1 & -1 & -2 & 3 & -2 \\
1 & -1 & -1 & 1 & 1 & -2 & -1 & 3 \\
1 & 1 & -1 & -1 & 1 & 2 & -1 & -3 \\
1 & 1 & -1 & -1 & -1 & 2 & 3 & 2 \\
1 & 2 & -1 & -2 & -1 & -1 & -1 & -1 \\
1 & 3 & 3 & 2 & 1 & 1 & -1 & 1 \\
\end{rsmallmatrix}
\right],
\end{align}
and
$
\mathbf{D}_1
=
\operatorname{diag}
\left(
\frac{1}{8},
\frac{1}{10},
\frac{1}{8},
\frac{1}{10},
\frac{1}{4},
\frac{1}{10},
\frac{1}{8},
\frac{1}{10}
\right)
$.
Considering the total energy error~\cite{Haweel2001SDCT,CB2012MRDCT}
between the exact and approximate matrices,
we obtained $3.32$ and $4.86$ 
as the error values
for the direct and inverse transformations,
respectively. 
Such errors are considered very small~\cite{Potluri2014Improved_Approx}.

Thus,
employing the orthogonalization procedure
described in~\cite{cintra2014dct_aprox},
we obtain the following expression for the DTT approximation:
$\hat{\mathbf{T}} = \mathbf{D}^\ast \cdot \mathbf{T}^\ast$,
where
$\mathbf{D}^\ast = 
\sqrt{
\operatorname{ediag}
\left(
\mathbf{T}^\ast
\cdot
(\mathbf{T}^\ast)^\top
\right)
}
=
\begin{bmatrix}
d_0^\ast & d_1^\ast & d_2^\ast & d_3^\ast & d_4^\ast & d_5^\ast & d_6^\ast & d_7^\ast
\end{bmatrix}^\top$
is a diagonal matrix
and
$\operatorname{ediag}(\cdot)$
returns 
a diagonal matrix
with the diagonal elements of its matrix argument~\cite{cintra2014dct_aprox}.
The inverse transformation is
$(\hat{\mathbf{T}})^{-1} 
=
(\mathbf{D}^\ast \cdot \mathbf{T}^\ast)^{-1}
=
(\mathbf{T}^\ast)^{-1} \cdot (\mathbf{D}^\ast)^{-1}
=
\mathbf{T}_1 \cdot \mathbf{D}_1 \cdot (\mathbf{D}^\ast)^{-1}$.
Therefore, 
the analysis and synthesis equations for the proposed transform
are given by
$\hat{\mathbf{X}}
=
\hat{\mathbf{T}}
\cdot
\mathbf{x}$
and
$\mathbf{x}
=
\mathbf{T}_1 \cdot \mathbf{D}_1 \cdot (\mathbf{D}^\ast)^{-1}
\cdot
\hat{\mathbf{X}}$,
where
$
\hat{\mathbf{X}}
=
\begin{bmatrix}\hat{X}_0 & \hat{X}_1 & \cdots & \hat{X}_7\end{bmatrix}^\top
$
is the approximate transformed vector.

\begin{figure}
\centering

\epsfig{file=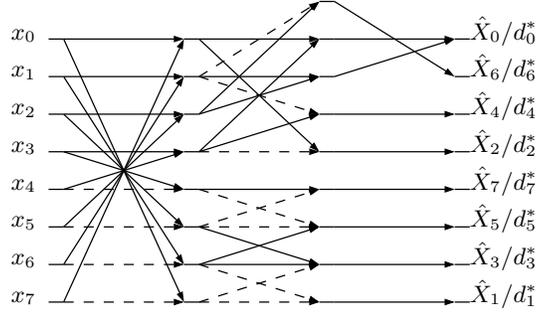}

\caption{%
Signal flow graph for 
$\mathbf{T}^\ast$.
Input data 
$x_n$, 
$n=0,1,\ldots,7$, 
relates to the output
$\hat{X}_k$, 
$k=0,1,\ldots,7$.
Dashed arrows
represent multiplications by
$-1$.
Scaling by $d_k^\ast$, $k=0,1,\ldots,7$,
can be ignored and absorbed into the quantization step.
}
\label{figure-fast-algorithm-1}
\end{figure}

However,
in several contexts,
diagonal matrices---such as $\mathbf{D}_1$ and $\mathbf{D}^\ast$---represent only scaling factors
and
may not contribute to
the computational cost of transformations.
For instance,
in JPEG-based image compression applications,
diagonal matrices can be embedded into
quantization block~\cite{CB2011RDCT,Swamy2013ITT,cintra2014dct_aprox}
and,
when the explicit transform coefficients are needless,
a scaled version of the transform-domain spectrum
is sufficient~\cite{arai1988fast_DCT}.
Therefore,
hereafter,
we disregard the diagonal matrices and
focus our analysis on the low-complexity matrices
$\mathbf{T}^\ast$ and $\mathbf{T}_1$.
A fast algorithm based
on sparse matrix factorization~\cite{BAS2008,CB2011RDCT,cintra2014dct_aprox}
was derived for the proposed 
forward and inverse approximations.
In Figure~\ref{figure-fast-algorithm-1},
the signal flow graph (SFG)
for the direct transformation is depicted.
The SFG for the inverse transformation
can be obtained according to the methods described
in~\cite{blahut_book}.
Moreover,
Table~\ref{table:complexity}
summarizes the arithmetic complexity assessment for the 
proposed transformations.
The fast algorithms for
$\mathbf{T}^\ast$
and
$\mathbf{T}_1$
demand
54.5\% and 34.1\% less additions than
the DTT fast algorithm (ITT)
proposed in~\cite{Swamy2013ITT}, 
respectively.

\begin{table}
\centering
\caption{
Arithmetic complexity of the proposed \mbox{1-D} transforms}
\label{table:complexity}
\begin{tabular}{lcccc}
\toprule

Method & Mult. & Additions & Shifts & Total
\\

\midrule
Exact DTT~\cite{Swamy2013ITT} & 
0 & 44 & 29 & 73
\\
Proposed $\mathbf{\hat{T}}^\ast$ & 
0 & 20 & 0 & 20
\\
Proposed $\mathbf{T}_1$& 
0 & 29 & 8 & 37
\\

\bottomrule
\end{tabular}
\end{table}

\section{Experimental Results}
\label{section-experimental}

\begin{figure}
\centering
\subfigure[SSIM]{\includegraphics[width=0.75\columnwidth]{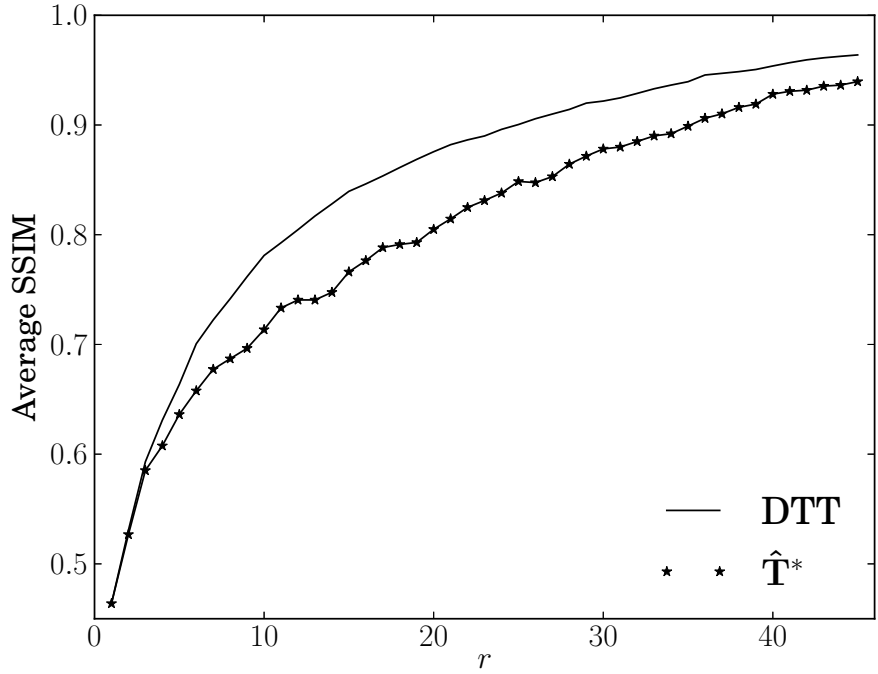}}
\subfigure[SR-SIM]{\includegraphics[width=0.75\columnwidth]{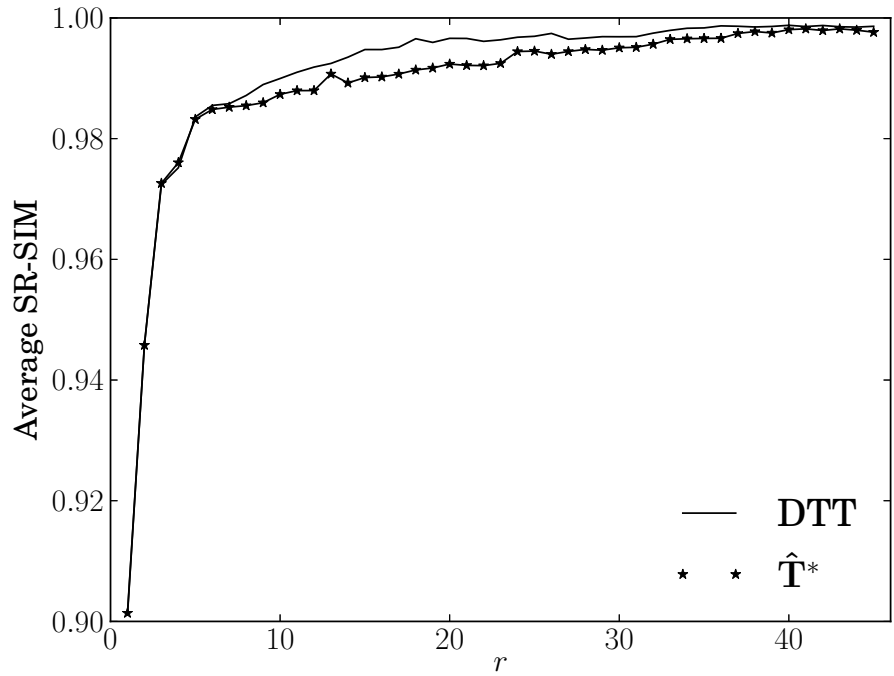}}
\caption{Quality metrics considering (a)~SSIM and (b)~SR-SIM
for the exact DTT and the proposed approximation in terms of~$r$.}
\label{figure:quality}
\end{figure}

\subsection{Image Compression}

\begin{figure}
\centering
\subfigure[DTT,~$r=6$]{\includegraphics[width=0.45\columnwidth]{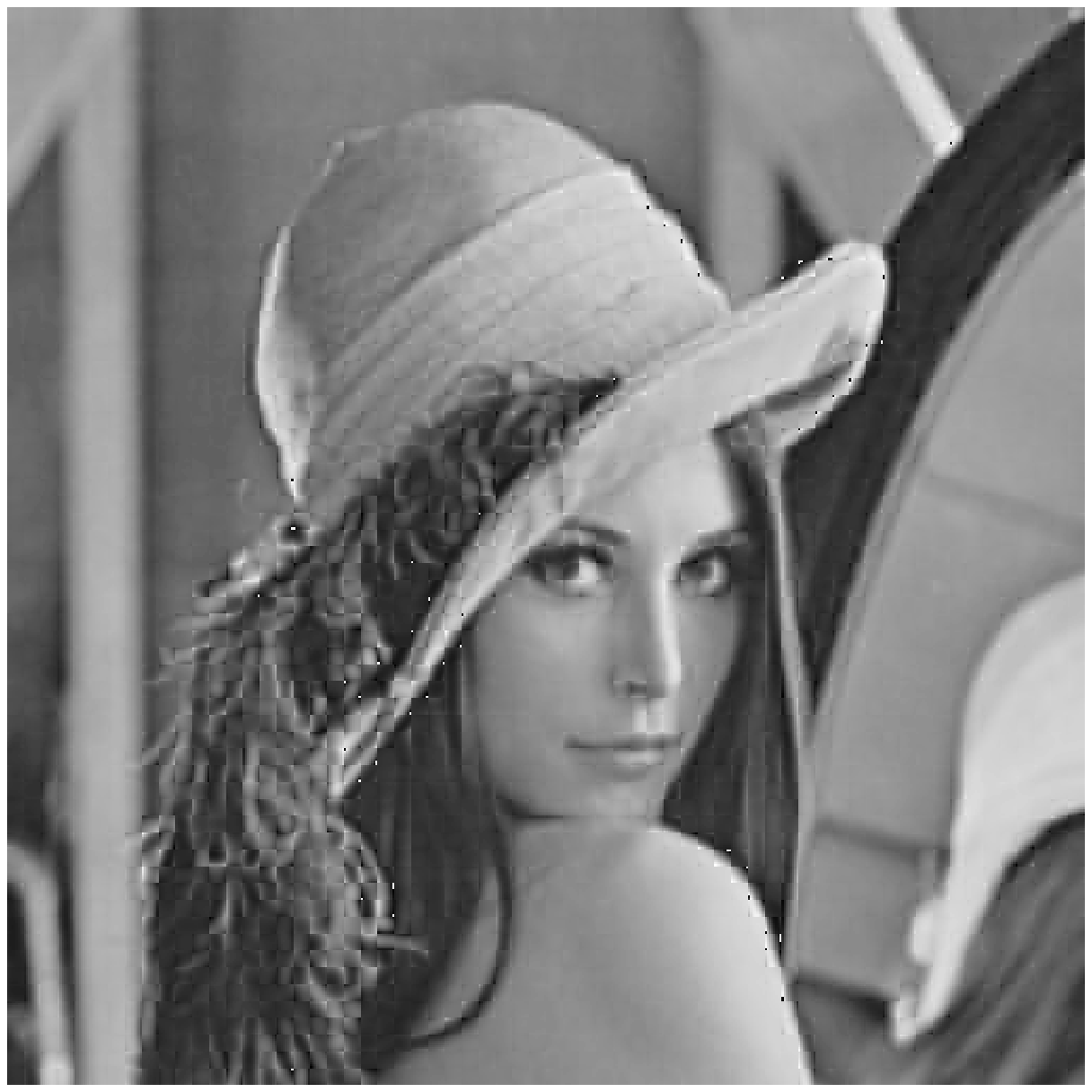}}
\subfigure[$\mathbf{\hat{T}}^\ast,r=6$]{\includegraphics[width=0.45\columnwidth]{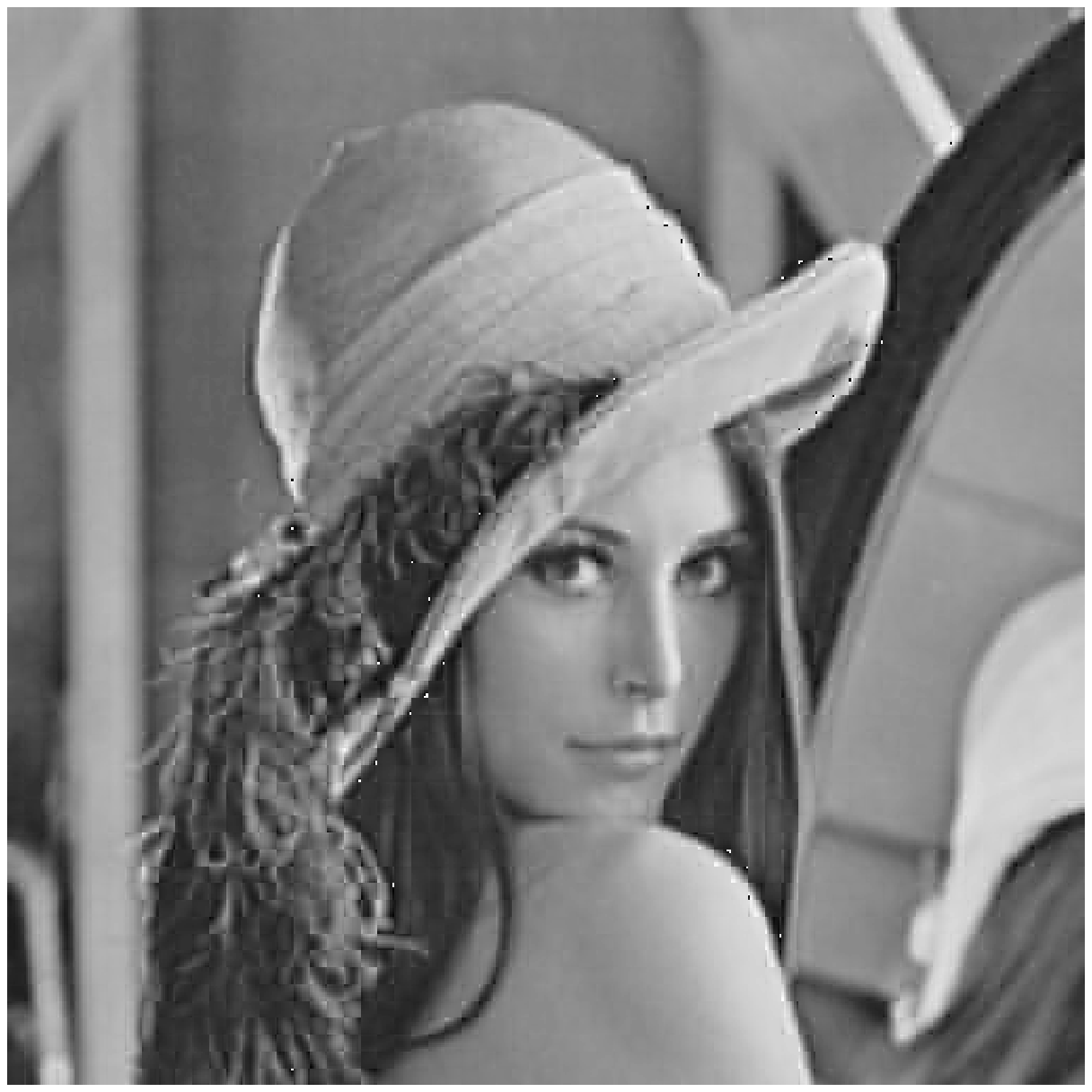}}
\caption{Compressed `Lena' image for $r = 6$ 
by means of the (a)~ DTT and (b)~the proposed approximation.}
\label{figure:lena}

\end{figure}

In order to assess the proposed transform 
in image compression applications,
we performed a JPEG-like simulation
based on~\cite{CB2011RDCT,Swamy2013ITT,cintra2014dct_aprox}.
A set of 45 512$\times$512 8-bit grayscale
images obtained from a standard public image bank~\cite{imagens} 
was considered.
Each image
was subdivided into 8$\times$8 size blocks
$\mathbf{A}_{i,j}$,
$i, j = 1, 2, \ldots, 64$.
Each block
is submitted 
to two-dimensional (\mbox{2-D}) versions of the discussed transformations
according to:
$\mathbf{B}_{i,j}
=
\mathbf{M}
\cdot
\mathbf{A}_{i,j}
\cdot
\mathbf{M}^\top$,
where $\mathbf{B}_{i,j}$ 
is the transform-domain block
and
$\mathbf{M} \in\{ \mathbf{T}, \mathbf{T}^\ast\}$
The resulting 64~spectral coefficients of each block were ordered
in the standard zigzag sequence.
Subsequently,
the $r$~initial coefficients in each block
were retained
and
the remaining coefficients were discarded~\cite{cintra2014dct_aprox}.
We adopted $1 \leq r \leq 45$.
Finally,
each transform-domain subimage
was submitted to inverse \mbox{2-D} transformations
and the full image was reconstructed.
Image quality measures
were employed to
assess
the
degradation
between original and reconstructed images.
The considered measures were
the structural similarity index (SSIM)~\cite{Bovik2004SSIM}
and the spectral residual base similarity (SR-SIM)~\cite{zhang2012SR_SIM}.
These measures have the distinction of 
being
consistent with
subjective ratings~\cite{Bovik2011SSIM,zhang2012SR_SIM}.
The peak signal-to-noise ratio (PSNR) 
was not considered as a figure of merit
because of its limited capability of
capturing
the
human perception of image fidelity and quality~\cite{bovik2009mse}.
For each value of $r$,
we considered
average measures across all considered images. 
Such methodology is
less prone to
variance effects and fortuitous data.
Figure~\ref{figure:quality} shows
the
resulting SSIM and SR-SIM measurements.
The proposed transform 
performed very closely to the exact DTT.
For qualitative purposes,
Figure~\ref{figure:lena} shows 
compressed images according to the DTT and the proposed approximation
for $r=6$;
images are visually indistinguishable.

\subsection{Video Compression}

\begin{figure}
\centering
\subfigure[]{\includegraphics[width=0.75\columnwidth]{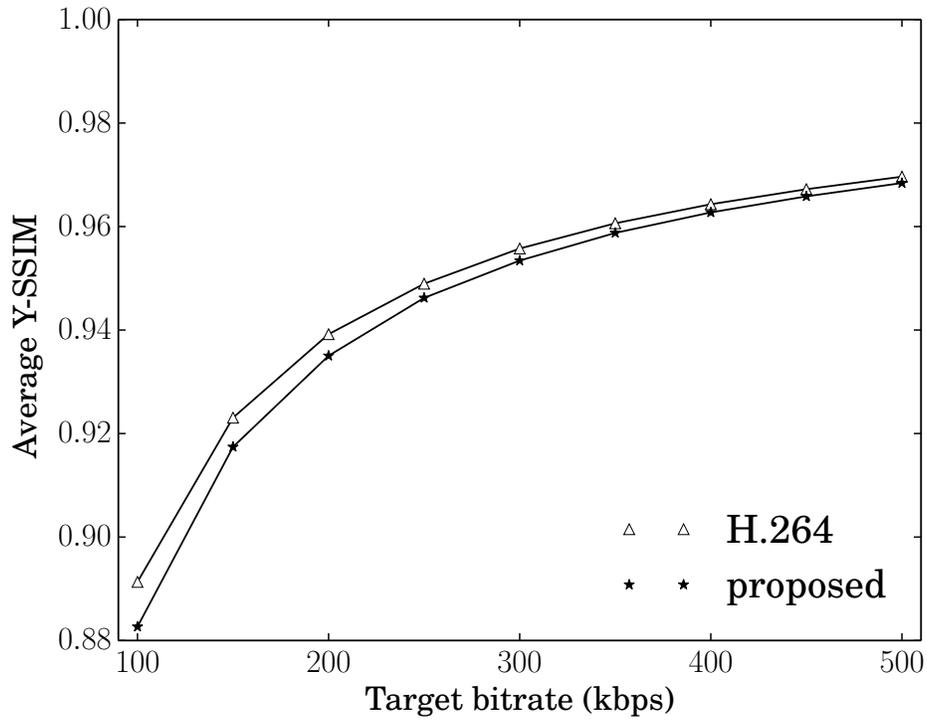}}
\subfigure[]{\includegraphics[width=0.75\columnwidth]{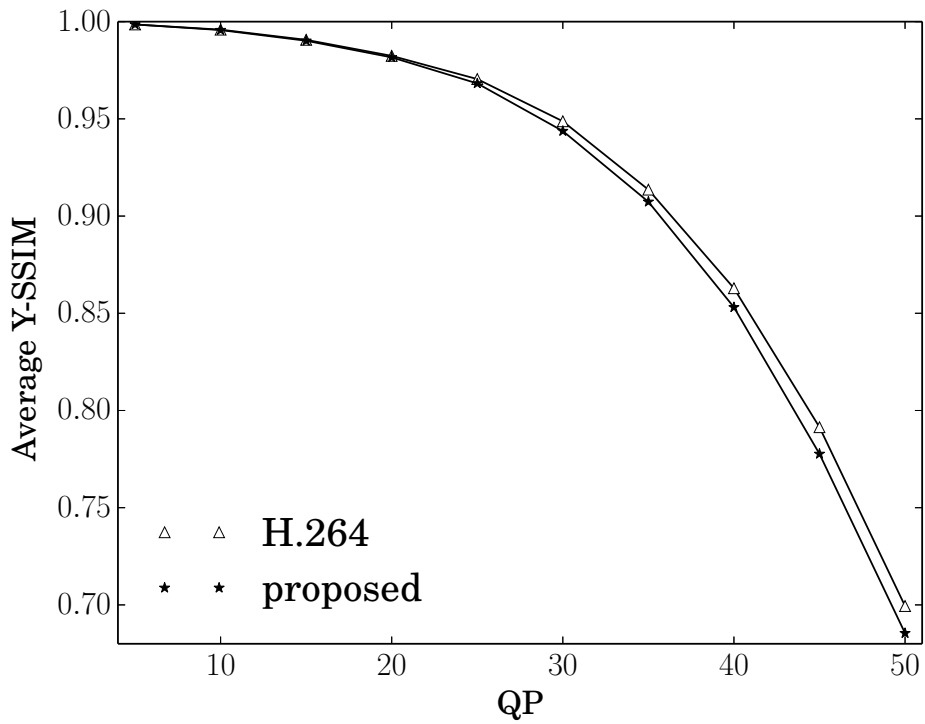}}
\caption{Video quality assessment in terms of 
(a)~fixed target bitrate and 
(b)~quantization parameter.}
\label{figure:bitrate_qp}
\end{figure}

\begin{figure}
\centering
\subfigure[H.264/AVC]{\includegraphics[width=0.45\columnwidth]{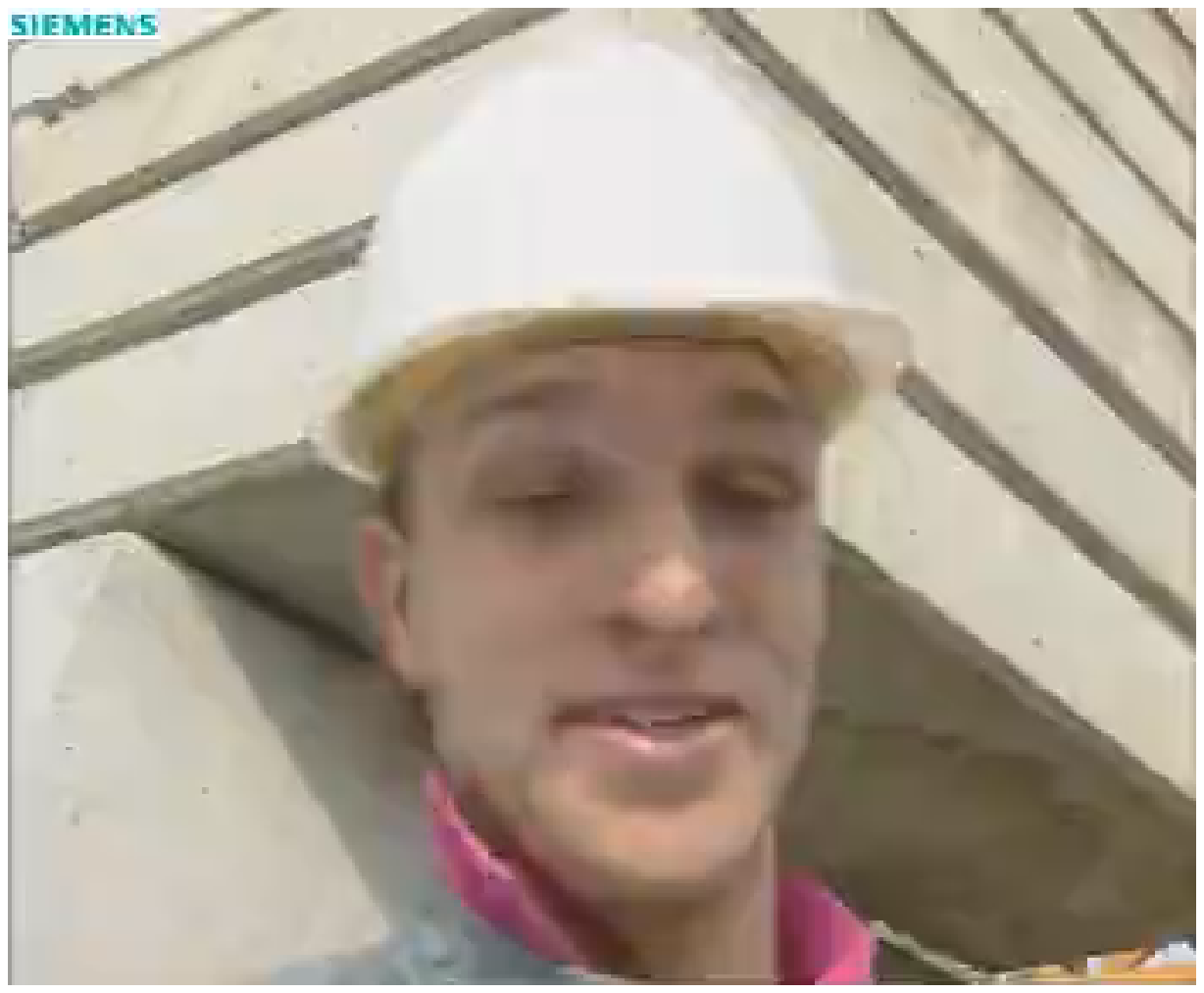}}
\subfigure[Modified H.264/AVC]{\includegraphics[width=0.45\columnwidth]{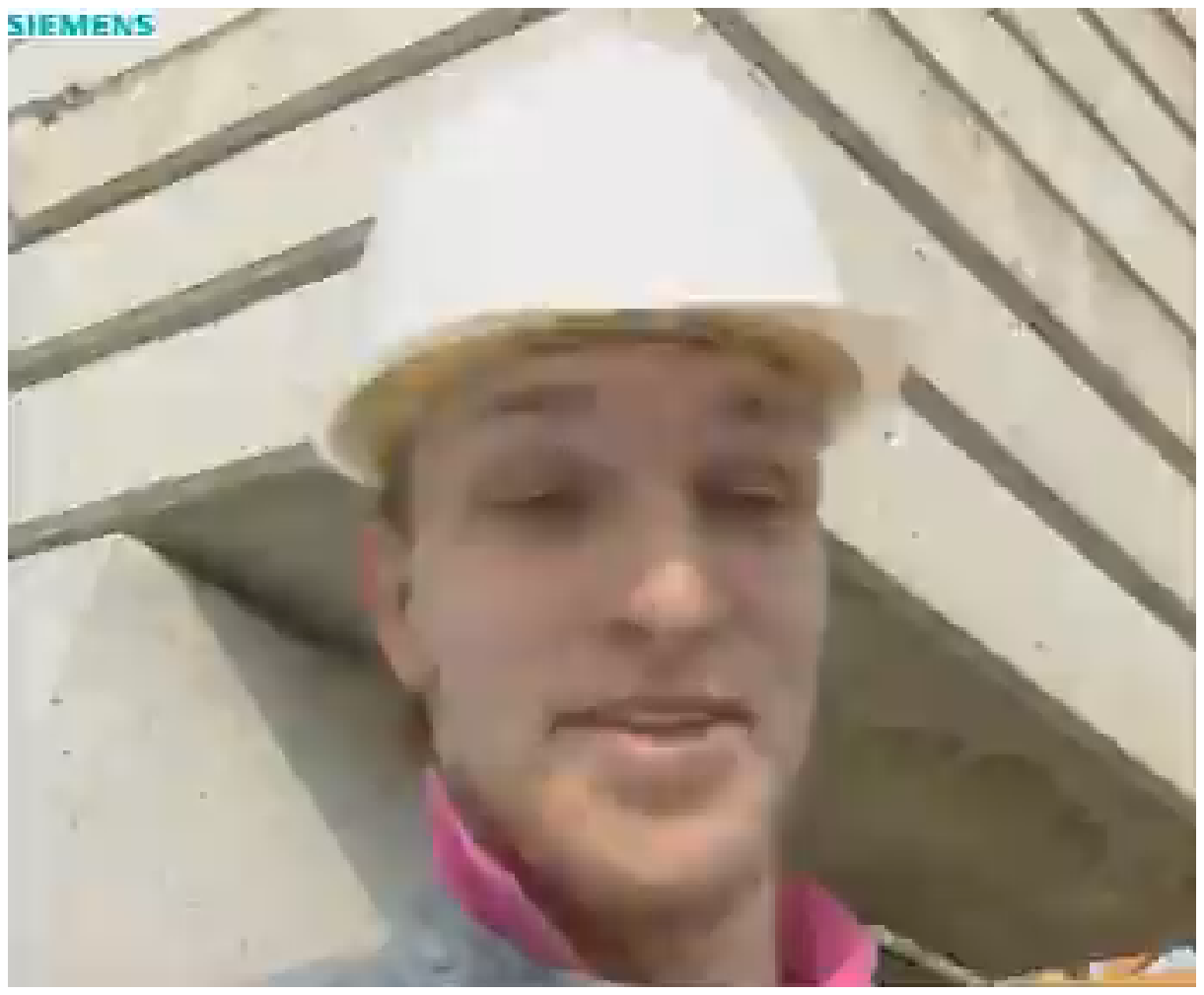}}

\caption{First frame of the compressed sequence
`Foreman'
according to
(a)~the original H.264/AVC
and
(b)~modified H.264/AVC with the proposed approximation.}

\label{figure:video_container}

\end{figure}

With the objective of assessing
the proposed transform performance in video coding, 
we have embedded the proposed DTT approximation 
in the widely employed
software library~x264~\cite{x264}
for encoding video streams into
the H.264/AVC standard~\cite{H264_book}.
The 8-point transform employed in H.264/AVC is
an integer approximation of the DCT
that
demands
32~additions and 14~bit-shifting operations~\cite{H264_8transform}.
In comparison,
the proposed 8-point direct transform
requires 38\%~less additions
and no bit-shifting operations,
while the proposed inverse transform
requires 9\%~less additions
and 43\%~less bit-shifting operations.
We encoded eleven CIF~videos 
with 
300~frames at 25~frames per second 
from a public video database~\cite{videos}
with the standard and the modified libraries.
In our simulation, 
we employed default settings and
controlled the video quality by two different approaches:
(i)~target bitrate, 
varying from 100 to 500~kbps with a step of 50~kbps
and
(ii)~quantization parameter~(QP), 
varying from 5~to 50~with steps of 5~units.
For video quality assessment,
we submitted the luma component
of the video frames
to 
average SSIM evaluation
relative to the Y~component (luminance).
Results are shown in Figure~\ref{figure:bitrate_qp}.
Even in scenarios of high compression (low bitrate/high QP),
the degradation related to
the proposed approximation 
is
in the order of 0.01~units of SSIM;
therefore, very low.
Figure~\ref{figure:video_container}
displays
 the first encoded frame of a standard video sequence
at low target bitrate (200~kbps).
The resulting compressed frames 
are visually 
indistinguishable.

\section{VLSI Architectures}\label{vlsi}

To compare hardware resource consumption of 
the proposed approximate DTT against the exact DTT proposed in~\cite{Swamy2013ITT}, 
the \mbox{1-D} version of both algorithms were
initially modeled and tested
in Matlab Simulink and
then were physically realized on 
a Xilinx Virtex-6 XC6VLX240T-1FFG1156 
field programmable gate array (FPGA)
device and 
validated using
hardware-in-the-loop testing through the JTAG interface. 
Both approximations were verified using more than 10000~test vectors
with complete agreement with theoretical values. 
Results are shown
in Table~\ref{fpgadtt}.
Metrics,
including 
configurable logic blocks (CLB) and flip-flop (FF)
count, 
critical path delay (CPD, in~ns), and
maximum operating frequency ($F_\text{max}$, in~MHz)
are provided.
In addition,
static ($Q_p$, in~mW) and 
frequency normalized
dynamic power ($D_p$, in~mW/MHz) consumptions
were
estimated using the Xilinx XPower Analyzer. 
The final throughput of the 1-D DTT was $438.68\times 10^6$
8-point
transformations/second, 
with a pixel rate of $3.509 \times 10^9$ pixels/second.
The percentage reduction in the number of CLBs and FFs
was 64.7\% and 71\%, 
respectively.
The dynamic power consumption $D_p$
of the proposed architecture was
44.9\% lower.
The figures of merit
area-time ($AT$) and 
area-time${}^2$ ($AT^2$) 
had percentage reductions of 66.1\% and 67.5\%
when compared with the exact DTT~\cite{Swamy2013ITT}. 

\begin{table}
\centering
\caption{Resource consumption on Xilinx XC6VLX240T-1FFG1156 de\-vice}
\label{fpgadtt}

\begin{tabular}{lcc}

\toprule
\multirow{2}{*}{Resource} & 
\multicolumn{2}{c}{Method} \\
\cmidrule{2-3}
& Exact DTT~\cite{Swamy2013ITT} & Proposed \\
\hline
CLB ($A$) & 408 & 144 \\
FF & 1370 & 396 \\
CPD ($T$) (ns) & 2.390 & 2.290 \\
$F_\text{max}$ (MHz)\! & 418.41 & 438.68 \\
$AT$ & 975.1 & 329.7 \\
$AT^2$ & 2330.5 & 755.1 \\
$D_p$ (mW/MHz)\! & 5.10 & 2.81 \\
$Q_p$ (W)\! & 3.44 & 3.44
\\
\bottomrule 
\end{tabular}

\end{table}

\section{Conclusion}
\label{conclusion}

In this paper,
a low-complexity approximation
for the 8-point DTT was proposed. 
The arithmetic cost of the proposed approximation
are significantly low,
when compared with the exact DTT.
At the same time,
the proposed tool
is very close to the DTT
in terms of image coding
for a wide range of compression rates.
In video compression,
the introduced approximation
was adapted into
the popular codec~H.264
furnishing virtually identical results
at a much less computational cost.
Our goal with the codec experimentation
is not to suggest the modification of an existing standard.
Our objective is to demonstrate the capabilities
of the proposed low-complexity transform 
in
asymmetric codecs~\cite{Mitra2004}. 
Such codecs are employed when a video is encoded once 
but decoded several times in 
low power devices~\cite{Mitra2004,Vijayanagar2014}.  
Additionally, 
the proposed transform can be 
considered
in distributed video coding (DVC)~\cite{Wyner1976DVC,Vijayanagar2014},
where
the computational complexity is concentrated in the decoder.
A relevant context
for DVC is
in remote sensors and video systems 
that are constrained in terms of power, bandwidth,
and computational capabilities~\cite{Vijayanagar2014}.
The proposed 
approximation 
is a viable alternative to the DTT;
possessing low-complexity and good performance
according to meaningful image quality measures.
Moreover,
the associated hardware realization
consumed roughly $1/3$ of the area required by the exact DTT;
also the dynamic power consumption 
was decreased by 
44.9\%.
Future work in this field may consider
the evaluation of DTT approximations
in quantization schemes~\cite{Ernawan2013quantization_tchebichef,Pratipati2013DTT_quant}.

\section*{Acknowledgments}  

This work was supported by
the 
CNPq, 
FACEPE,
and FAPERGS, 
Brazil;
and 
the University of Akron, Ohio, USA.

\bibliographystyle{ieeetr}
\bibliography{adtt_ref} 

\begin{thebibliography}{10}

\bibitem{Mukundan2001DTM}
R.~Mukundan, S.~Ong, and P.~A. Lee, ``Image analysis by {T}chebichef moments,''
  {\em IEEE Transactions on Image Processing}, vol.~10, no.~9, pp.~1357--1364,
  2001.

\bibitem{Leida2014artifact_tchebichef}
L.~Leida, Z.~Hancheng, Y.~Gaobo, and Q.~Jiansheng, ``Referenceless measure of
  blocking artifacts by {T}chebichef kernel analysis,'' {\em IEEE Signal
  Processing Letters}, vol.~21, pp.~122--125, Jan 2014.

\bibitem{Roux2012blind}
H.~Huang, G.~Coatrieux, H.~Shu, L.~Luo, and C.~Roux, ``Blind integrity
  verification of medical images,'' {\em IEEE Transactions on Information
  Technology in Biomedicine}, vol.~16, pp.~1122--1126, Nov 2012.

\bibitem{Ernawan2013quantization_tchebichef}
F.~Ernawan, N.~Abu, and N.~Suryana, ``{TMT} quantization table generation based
  on psychovisual threshold for image compression,'' in {\em 2013 International
  Conference of Information and Communication Technology (ICoICT)},
  pp.~202--207, Mar 2013.

\bibitem{Pratipati2013DTT_quant}
S.~Prattipati, M.~Swamy, and P.~Meher, ``A variable quantization technique for
  image compression using integer tchebichef transform,'' in {\em 2013 9th
  International Conference on Information, Communications and Signal Processing
  (ICICS)}, pp.~1--5, Dec 2013.

\bibitem{Swamy2013ITT}
S.~Prattipati, S.~Ishwar, P.~Meher, and M.~Swamy, ``A fast 8$\times$8 integer
  {T}chebichef transform and comparison with integer cosine transform for image
  compression,'' in {\em 2013 IEEE 56th International Midwest Symposium on
  Circuits and Systems (MWSCAS)}, pp.~1294--1297, 2013.

\bibitem{Senapati2014listlessDTT}
R.~Senapati, U.~Pati, and K.~Mahapatra, ``Reduced memory, low complexity
  embedded image compression algorithm using hierarchical listless discrete
  {T}chebichef transform,'' {\em IET Image Processing}, vol.~8, pp.~213--238,
  Apr 2014.

\bibitem{Li2008WSN}
L.~W. Chew, L.-M. Ang, and K.~P. Seng, ``Survey of image compression algorithms
  in wireless sensor networks,'' in {\em 2008 International Symposium on
  Information Technology (ITSim)}, vol.~4, pp.~1--9, Aug 2008.

\bibitem{Ernawan2011mobile_tchebichef}
F.~Ernawan, E.~Noersasongko, and N.~Abu, ``An efficient 2$\times$2 {T}chebichef
  moments for mobile image compression,'' in {\em 2011 International Symposium
  on Intelligent Signal Processing and Communications Systems (ISPACS)},
  pp.~1--5, Dec 2011.

\bibitem{Kouadria2013WSN}
N.~Kouadria, N.~Doghmane, D.~Messadeg, and S.~Harize, ``Low complexity {DCT}
  for image compression in wireless visual sensor networks,'' {\em Electronics
  Letters}, vol.~49, pp.~1531--1532, Nov 2013.

\bibitem{CB2011RDCT}
R.~J. Cintra and F.~M. Bayer, ``A {DCT} approximation for image compression,''
  {\em IEEE Signal Processing Letters}, vol.~18, pp.~579--582, Oct 2011.

\bibitem{cintra2014dct_aprox}
R.~J. Cintra, F.~M. Bayer, and C.~J. Tablada, ``Low-complexity 8-point {DCT}
  approximations based on integer functions,'' {\em Signal Processing},
  vol.~99, pp.~201--214, 2014.

\bibitem{Haweel2001SDCT}
T.~I. Haweel, ``A new square wave transform based on the {DCT},'' {\em Signal
  Processing}, vol.~81, no.~11, pp.~2309--2319, 2001.

\bibitem{BAS2008}
S.~Bouguezel, M.~Ahmad, and M.~Swamy, ``A multiplication-free transform for
  image compression,'' in {\em 2008 2nd International Conference on Signals,
  Circuits and Systems (SCS)}, pp.~1--4, Nov 2008.

\bibitem{BAS2011}
S.~Bouguezel, M.~Ahmad, and M.~Swamy, ``A low-complexity parametric transform
  for image compression,'' in {\em 2011 IEEE International Symposium on
  Circuits and Systems (ISCAS)}, pp.~2145--2148, May 2011.

\bibitem{BAS2013}
S.~Bouguezel, M.~Ahmad, and M.~Swamy, ``Binary discrete cosine and {H}artley
  transforms,'' {\em IEEE Transactions on Circuits and Systems I: Regular
  Papers}, vol.~60, pp.~989--1002, Apr 2013.

\bibitem{LO2004Dct_appr}
K.~Lengwehasatit and A.~Ortega, ``Scalable variable complexity approximate
  forward {DCT},'' {\em IEEE Transactions on Circuits and Systems for Video
  Technology}, vol.~14, pp.~1236--1248, Nov 2004.

\bibitem{CB2012MRDCT}
F.~M. Bayer and R.~J. Cintra, ``{DCT}-like transform for image compression
  requires 14 additions only,'' {\em Electronics Letters}, vol.~48,
  pp.~919--921, Jul 2012.

\bibitem{Potluri2014Improved_Approx}
U.~S. Potluri, A.~Madanayake, R.~J. Cintra, F.~M. Bayer, S.~Kulasekera, and
  A.~Edirisuriya, ``Improved 8-point approximate {DCT} for image and video
  compression requiring only 14 additions,'' {\em IEEE Transactions on Circuits
  and Systems I: Regular Papers}, vol.~61, pp.~1727--1740, Jun 2014.

\bibitem{Wallace1992JPEG}
G.~Wallace, ``The {JPEG} still picture compression standard,'' {\em IEEE
  Transactions on Consumer Electronics}, vol.~38, pp.~xviii--xxxiv, Feb 1992.

\bibitem{H264_book}
I.~Richardson, {\em The {H}.264 Advanced Video Compression Standard}.
\newblock John Wiley and Sons, 2~ed., 2010.

\bibitem{Sullivan2012HEVC}
G.~Sullivan, J.~Ohm, W.-J. Han, and T.~Wiegand, ``Overview of the high
  efficiency video coding ({HEVC}) standard,'' {\em IEEE Transactions on
  Circuits and Systems for Video Technology}, vol.~22, no.~12, pp.~1649--1668,
  2012.

\bibitem{HTF53}
H.~Bateman, A.~Erd{\'e}lyi, W.~Magnus, F.~Oberhettinger, and F.~Tricomi, {\em
  Higher transcendental functions}, vol.~2.
\newblock McGraw-Hill, 1953.

\bibitem{Gray1998Quant}
R.~Gray and D.~Neuhoff, ``Quantization,'' {\em IEEE Transactions on Information
  Theory}, vol.~44, pp.~2325--2383, Oct 1998.

\bibitem{arai1988fast_DCT}
Y.~Arai, T.~Agui, and M.~Nakajima, ``A fast {DCT-SQ} scheme for images,'' {\em
  IEICE Transactions}, vol.~E71, pp.~1095--1097, Nov 1988.

\bibitem{blahut_book}
R.~Blahut, {\em Fast Algorithms for Signal Processing}.
\newblock Cambridge University Press, 2010.

\bibitem{imagens}
{University of Southern California, Signal and Image Processing Institute},
  ``The {USC-SIPI} image database.'' \url{http://sipi.usc.edu/database/}, 2014.

\bibitem{Bovik2004SSIM}
Z.~Wang, A.~Bovik, H.~Sheikh, and E.~Simoncelli, ``Image quality assessment:
  from error visibility to structural similarity,'' {\em IEEE Transactions on
  Image Processing}, vol.~13, pp.~600--612, Apr 2004.

\bibitem{zhang2012SR_SIM}
L.~Zhang and H.~Li, ``{SR-SIM}: A fast and high performance {IQA} index based
  on spectral residual,'' in {\em 19th IEEE International Conference on Image
  Processing (ICIP)}, pp.~1473--1476, Sep 2012.

\bibitem{Bovik2011SSIM}
Z.~Wang and A.~Bovik, ``Reduced- and no-reference image quality assessment,''
  {\em IEEE Signal Processing Magazine}, vol.~28, pp.~29--40, Nov 2011.

\bibitem{bovik2009mse}
Z.~Wang and A.~Bovik, ``Mean squared error: Love it or leave it? a new look at
  signal fidelity measures,'' {\em IEEE Signal Processing Magazine}, vol.~26,
  pp.~98--117, Jan 2009.

\bibitem{x264}
x264 team, ``x264.'' \url{http://www.videolan.org/developers/x264.html}, 2014.

\bibitem{H264_8transform}
S.~Gordon, D.~Marpe, and T.~Wiegand, ``Simpliﬁed use of 8$\times$8 transform
  -- updated proposal and results.'' Joint Video Team ({JVT}) of {ISO/IEC MPEG}
  and {ITU-T VCEG}, doc. {JVT}--{K028}, Munich, Germany, Mar 2004.

\bibitem{videos}
``{Xiph.org Video Test Media}.'' \url{https://media.xiph.org/video/derf/},
  2014.

\bibitem{Mitra2004}
U.~Mitra, {\em Introduction to Multimedia Systems}.
\newblock Academic Press, 2004.
\newblock 207 p.

\bibitem{Vijayanagar2014}
K.~R. Vijayanagar, J.~Kim, Y.~Lee, and J.~bok Kim, ``Low complexity distributed
  video coding,'' {\em Journal of Visual Communication and Image
  Representation}, vol.~25, no.~2, pp.~361--372, 2014.

\bibitem{Wyner1976DVC}
A.~Wyner and J.~Ziv, ``The rate-distortion function for source coding with side
  information at the decoder,'' {\em IEEE Transactions on Information Theory},
  vol.~22, pp.~1--10, Jan 1976.

\end{thebibliography}

\end{document}